# Validation of an immersed boundary method for compressible flows

T. Renaud[1], C. Benoit[1], S. Péron[1], I. Mary[1] and N. Alferez[1]
*ONERA – Université Paris Saclay, F-92322, Châtillon, France*

**This paper sums up some recent validations of an immersed boundary method for compressible flow simulations. It has been already shown that this method is able to provide accurate results without meshing effort around more or less complex geometries. Here, the authors focus on several simple test cases to assess the application range of the method. From subsonic to hypersonic flows, from steady to large-eddy unsteady simulations, in 2D or 3D, results are compared to classical simulations with body-fitted meshes and experiment. Moreover, the use of an efficient Navier-Stokes solver for all the cases is part of the benefit of the approach to provide a CFD solution in a reduced time.**

## I. Nomenclature

| | | |
|---|---|---|
| *IBM* | = | immersed boundary method |
| *IBC* | = | immersed boundary conditions |
| *d* | = | cylinder diameter |
| $C_p$ | = | pressure coefficient |
| c | = | chord |
| $M_\infty$ | = | Mach number |
| Re | = | Reynolds number |
| α | = | angle of attack |
| $\eta_1$ | = | wall adiabatic efficiency |

## II. Introduction

In the past decades, a wide range of immersed boundary approaches have been developed to deal with incompressible flows for elastic and moving obstacles Due to the increase of the geometrical complexity of configurations studies in aeronautics in the last decade, many researchers have developed immersed boundary approaches to solve compressible flows [6][7]. An immersed boundary method (IBM) has been developed for five years at ONERA [13] in combination with an automatic adaptive Cartesian mesh generation to deal with complex geometries. This paper focuses on the validation of the immersed boundary method for a wide range of flow regimes and Mach numbers.

First, a short description of the current immersed boundary method and of the flow solver is provided. Then, inviscid and viscous flow simulations around a NACA0012 profile using the IBM on a set of Cartesian grids are performed and compared to a classical structured multi-block approach. A grid convergence study is achieved.
Then, a test-case where the immersed boundary conditions mimic both a wall boundary and an injection is presented. Then, the simulation of the unsteady flow around a cylinder at Reynolds number of 3900 is performed. Results are compared with the experiment and discussed. Finally, the simulation of a supersonic flow is presented.

.

[1] Research scientist, Department of Aerodynamics, Aeroelasticity & Acoustics (DAAA).

## III. Description of the immersed boundary method

This paper focuses on the validation of an immersed boundary approach for compressible flow simulations. The main advantage of this kind of approach is that the mesh does not need to conform to the obstacles, consequently reducing the setup time of a numerical simulation.

Numerous approaches exist in the literature, as reviewed by Mittal & Iaccarino [24], among which continuous forcing approaches, as developed by Peskin [8][9] and discrete forcing approaches. Our approach consists of a Ghost Cell direct forcing approach, similar to the one of Fadlun et al. [11] and Ferziger et al. [12].

In our approach, an octree-based Cartesian mesh is automatically generated [15] around the set of obstacles defined by surface meshes. This enables to take fully advantage of the immersed boundary method, since the mesh is automatically generated and is able to adapt to geometrical details and flow features, as it will be demonstrated in paragraph V.D. Solid points are identified by a blanking technique, usually applied for overset grid methods [16]. Target points in the vicinity of solid points are marked. These target points are those where the solution will be forced at each time step to mimic a boundary condition. A preprocessing step is required to compute all the information needed for the reconstruction of the solution at each time step for these points. The algorithm is detailed in [13]. Since the IBM is applied on Cartesian grid cells, a wall model must be applied for viscous flow simulations to prevent a huge number of mesh points, due to the cell size requirements near the wall. In this paper, Musker's algebraic wall model is used [23].

Recently, we have extended the method to be able to deal with other immersed boundary conditions, such as injection or imposed pressure. This enables new applications to be tackled, such as the one showed in paragraph V.B.

The IBM pre and post-processing are achieved using Cassiopee package [16].

## IV. Description of the solver

A Navier-Stokes solver named FAST (Flexible Aerodynamic Solver Technology) has been developed at ONERA for three years [21], whose objective is to provide a software architecture and numerical techniques allowing for a high level of efficiency, flexibility and upgradeability. This demonstrator is made by a set of independent modules, each of them defining a CFD solver dedicated to Cartesian, curvilinear and polyhedral grids, which rely on the CGNS/Python data representation. These modules work in an interoperable way with other modules, most of them being Cassiopee modules devoted to pre, co and post-processing.

Each CFD solver solves the compressible Euler and Navier-Stokes equations using a second-order accurate Finite-Volume Method only for interior points of each grid. The turbulence modeling is handled by the Spalart-Allmaras model in the RANS mode, whereas a MILES approach is retained for LES [22]. For RANS computations, the Roe-MUSCL scheme is used with an implicit time integration and a local time stepping. Jacobian approximations are those proposed by Jameson & Yoon and Coakley, whereas the linear system is solved by the LU-SGS method. For LES computations, an hybrid centered/upwind scheme is retained to manage a good compromise between robustness and accurate simulation of the turbulent small eddies, whereas the temporal integration is achieved by a three-step Runge-Kutta explicit scheme, or by a second-order implicit Gear scheme with local Newton sub-iterations.

Moreover, the FAST solver is optimized for regular Cartesian grids, simplifying flux formula and metrics, gaining both in CPU time and memory. Besides, it has a highly efficient parallelization based on hybrid MPI/OpenMP approach. The computational load is distributed between blocks and under-blocks, enabling cache blocking. The solver has been optimized within an Intel partnership, called IPCC. The CPU cost for the following test-cases is about 0.4µs/points/iteration on Haswell core (Intel® Xeon® CPU E5-2690 v3 @ 2.60GHz). The FAST IBM approach does not require a lot of memory and, thus, for example, a wing-body configuration with 690M points can be easily simulated on 256 cores [13].

## V. Validation

The objective is to validate the present IBM approach and to demonstrate its capability to simulate a wide range of test-cases: 2D/3D configurations, Euler/Navier-Stokes turbulent/LES equations, steady/unsteady simulations, subsonic/transonic/supersonic flows… For each case, a comparison with experimental or computational results using a classical structured multi-block approach will quantify the accuracy of the method.

### V.A. Grid convergence for subsonic and transonic flows around the NACA0012 profile

The objective of this part is to demonstrate the capability of the IBM approach to converge in terms of mesh cells size. For that purpose, two well-known test-cases are here proposed around the NACA0012 profile. The Cartesian grid is automatically performed with a uniform cell size all around the profile. The domain size corresponds to 500c (where c is the profile chord) in each direction. The mesh is refined progressively by dividing this cell size by two to obtain the grid convergence. Table 1 gives the features of the different meshes used here. The body-fitted grid and n example of the Cartesian grid around the profile are shown on Figure 1. The results of the IBM method are compared to a classical CFD simulation done with the same solver FAST but a body-fitted grid. This body-fitted mesh is refined near the wall in order to capture the boundary layer for the turbulent test-case, where the Spalart-Allmaras turbulence model is applied. It is a very fine grid as 3000 points discretize the upper side of the profile and 1000 points are put in the normal direction. In this case, the IBM meshes are coarser than the body-fitted one but generally the IBM approach consumes more points than a body-fitted mesh method due to the Cartesian structure of the grid.

**Table 1 – Mesh sizes around the NACA0012 profile**

| | Body-fitted | IBM1 | IBM2 | IBM3 | IBM4 | IBM5 | IBM6 |
|---|---|---|---|---|---|---|---|
| Near-body cell size (*chord) | $3.\ 10^{-6}$ (normal to the wall) | $5.\ 10^{-3}$ | $2.5\ 10^{-3}$ | $1.25\ 10^{-3}$ | $6.\ 10^{-4}$ | $3.\ 10^{-4}$ | $1.5\ 10^{-4}$ |
| Number of 2D points (*$10^6$ points) | 8. | 0.25 | 0.31 | 0.42 | 0.6 | 0.95 | 1.96 |
| Number of blocks | 4 | 87 | 95 | 103 | 129 | 166 | 270 |

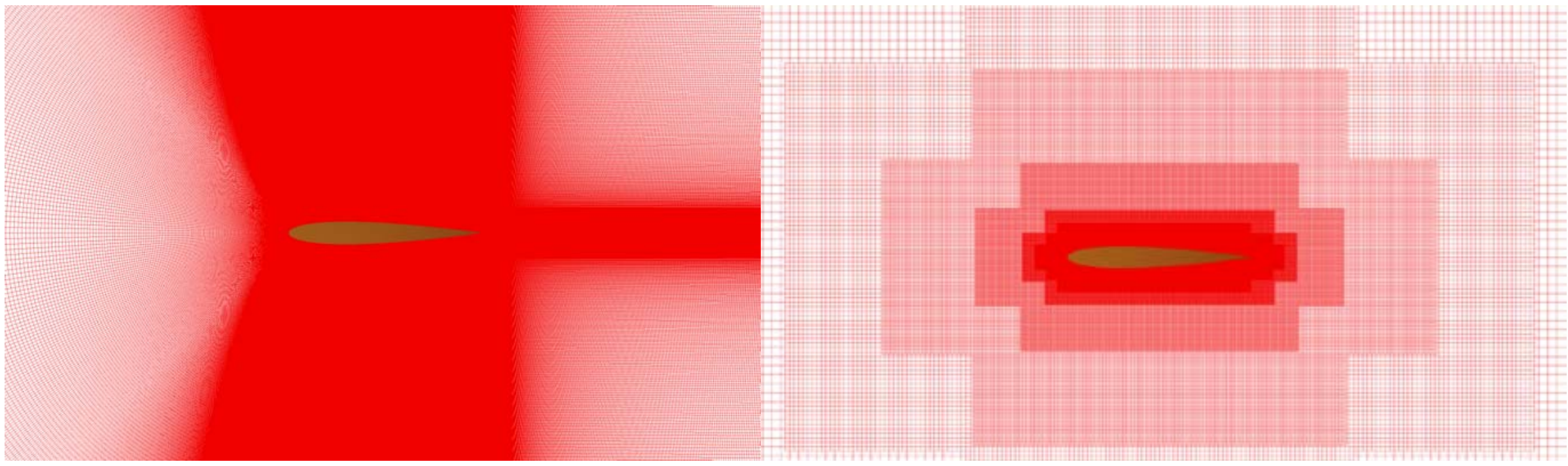

**Figure 1 – Body-fitted (left) and IBM (right) meshes around the NACA0012 profile**

The first test-case is the 2D simulation of the NACA0012 profile for the Euler equations in a transonic flow. The Mach number is $M_\infty$=0.85 for an angle of attack α=1°.

The flow solutions obtained with the body-fitted and IBM6 grids are shown on Figure 2 with the iso-contours of Mach number. The results are in good agreement, in particular near the profile. While going away from the near-body region, the coarser levels of the Cartesian grid tend to dissipate the shocks. Figure 3 presents the grid convergence in terms of pressure coefficient on the wall surface. The main differences between the grids are the capture of the shock locations and the pressure gradient on the leading and trailing edges. The IBM approach shows a good grid convergence, as the finest grid results matches the body-fitted grid ones. On the upper surface, the shock captured by the IBM approach is located upstream the body-fitted reference solution by 0.003c. On the lower surface, the shock captured by the IBM approach is located downstream the body-fitted reference solution by 0.01c.

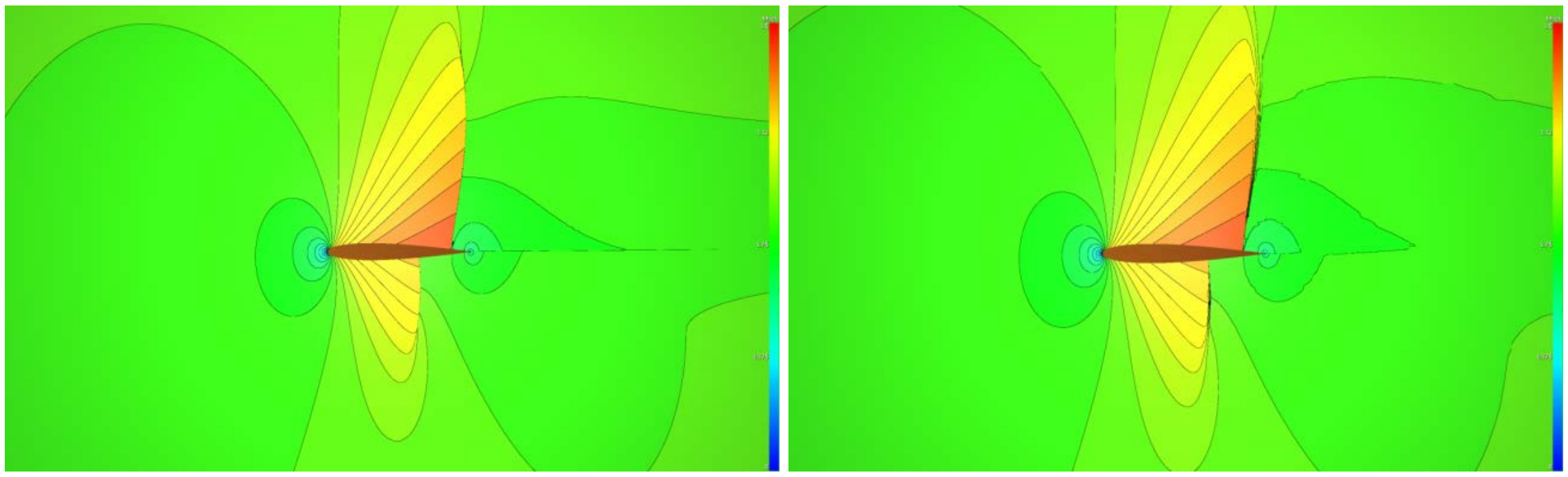

**Figure 2 – Mach number flowfield around the NACA0012 profile at $M_\infty$=0.85 and α=1° - Comparison between the body-fitted mesh results (left) and the IBM6 mesh results (right)**

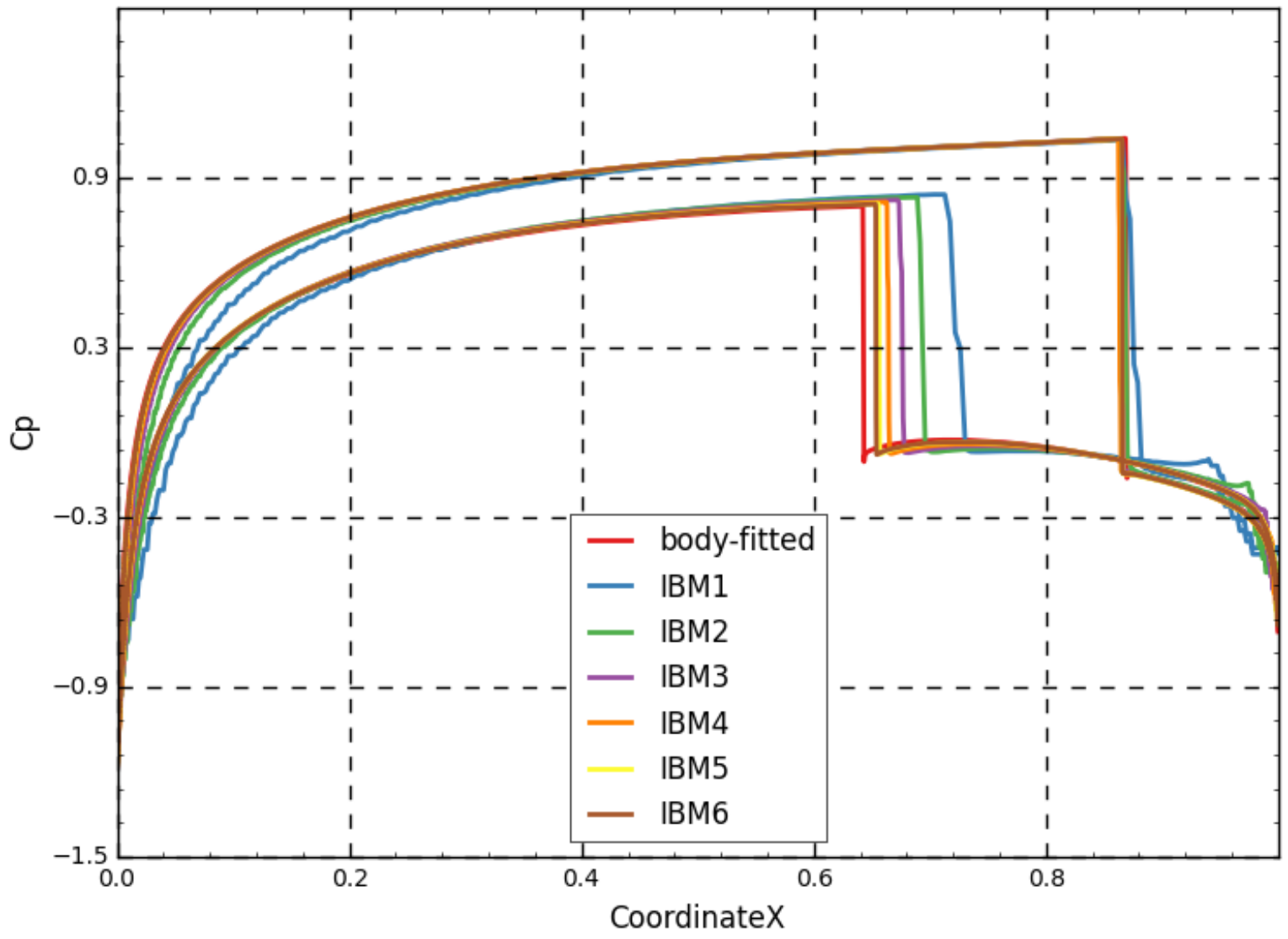

**Figure 3 - Grid convergence for the pressure coefficient around the NACA0012 profile at $M_\infty$=0.85 and α=1°**

The second test-case is the 2D simulation of the NACA0012 profile for the Navier-Stokes turbulent equations in a subsonic flow. The Mach number is $M_\infty$=0.15 for an angle of attack α=0° and the Reynolds number is Re=6 $10^6$.

The flow solutions obtained with the body-fitted and IBM5 grids are shown on Figure 4 with the iso-contours of Mach number. The results are in good agreement. The coarser levels of the Cartesian grid tend to dissipate the wake of the profile. For the body-fitted results, the wake is better captured due to the C-mesh topology of the grid. Figure 5 presents the grid convergence in terms of pressure coefficient on the wall surface. The main differences between the grids are located on the profile leading edge. The IBM approach shows a good grid convergence for the pressure coefficient, as the finest grid results (IBM6) matches the body-fitted grid ones. Figure 6 shows the grid convergence in terms of friction coefficient. Even if the results get closer to the reference body-fitted solution with the grid resolution, the actual IBM method is not able to predict correctly the friction coefficient. It is more due to the present wall law, as described in [13] (the same wall law applied on a coarse body-fitted grid would exhibit the same discrepancy for the friction coefficient).

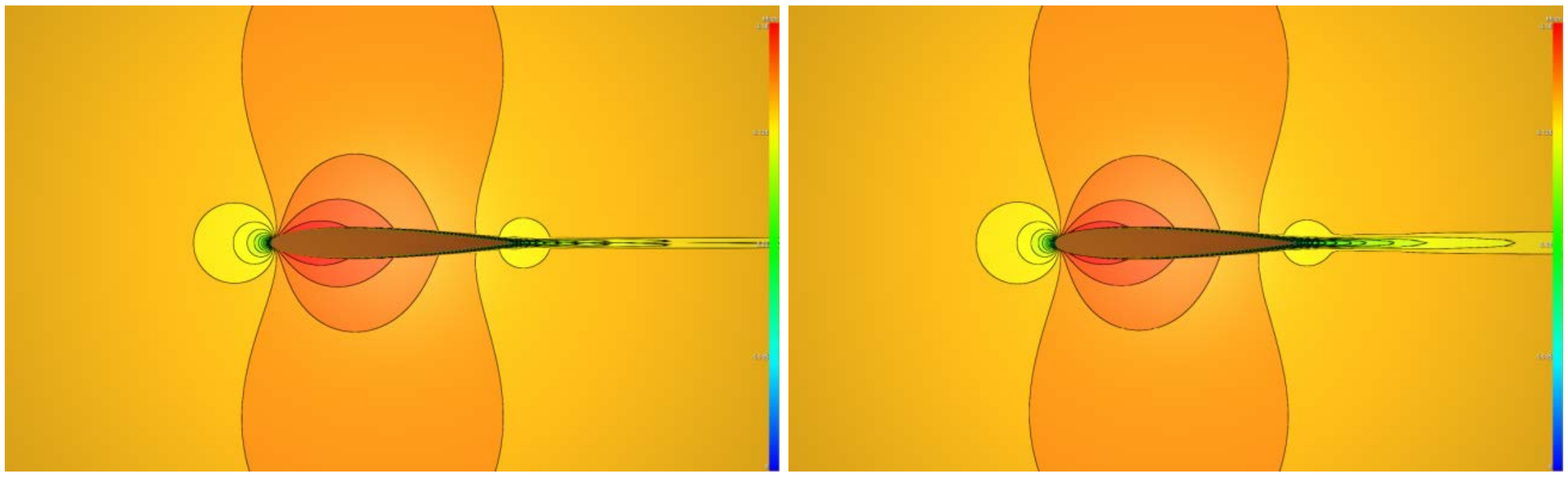

**Figure 4 - Mach number flowfield around the NACA0012 profile at $M_\infty$=0.15 and α=0° - Comparison between the body-fitted mesh results (left) and the IBM5 mesh results (right)**

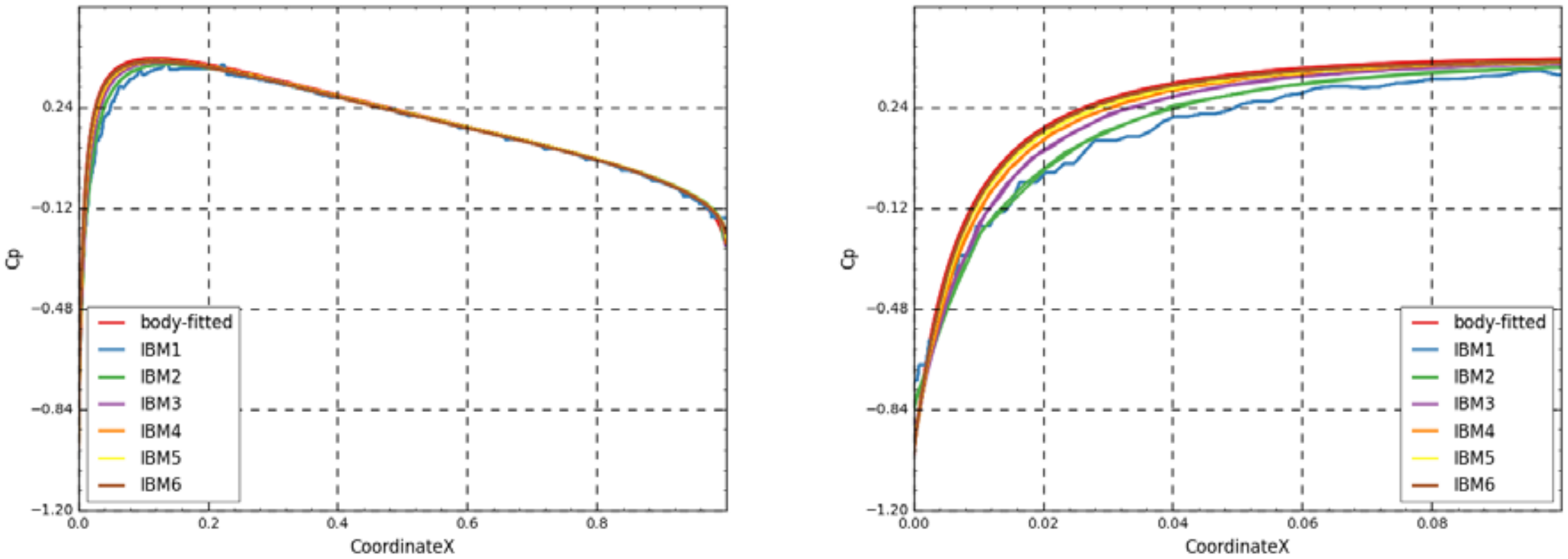

**Figure 5 - Grid convergence for the pressure coefficient around the NACA0012 profile at $M_\infty$=0.15 and α=0°**

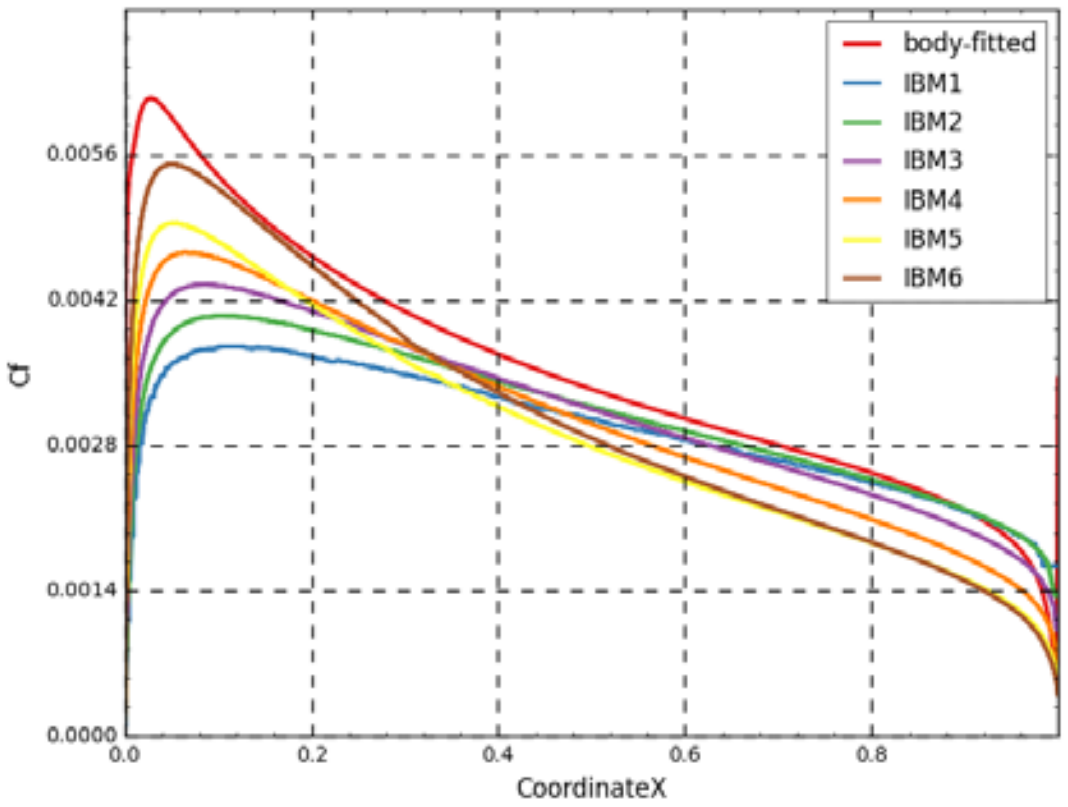

**Figure 6 - Grid convergence for the friction coefficient around the NACA0012 profile at $M_\infty$=0.15 and α=0°**

### V.B. 3D hot jet interaction with a wall in a cross flow

In 2004, one of the objectives of the MAEVA project [1] performed at ONERA in cooperation with Airbus and Cerfacs was to study the interaction between a hot jet and a cross-flow, which is a typical interaction that occurs when the exhaust hot jet of the engine used for de-icing impinges the nacelle, leading to an undesirable local overheating. For that purpose, wind tunnel tests were performed in the first phase of MAEVA in the ONERA F2 wind tunnel on a nacelle profile equipped with hot exhaust jet. Figure 7 presents the wind tunnel setup and a slice of the geometry shows the internal cavity shape leading to a square hole on the profile surface. For this test-case, the Mach number is $M_\infty$=0.138 for an angle of attack $\alpha$=1°. The Reynolds number is Re=3.1 $10^6$ and the hot jet temperature and mass-flow conditions are $T_{inj}$=80°C, $\rho_{inj}$=17.7g/s. During this project, computations have been performed with the elsA solver [20], developed by ONERA for industrial purpose, on classical body-fitted structured meshes with different turbulence models (k-l, Durbin k-ε v²f …).

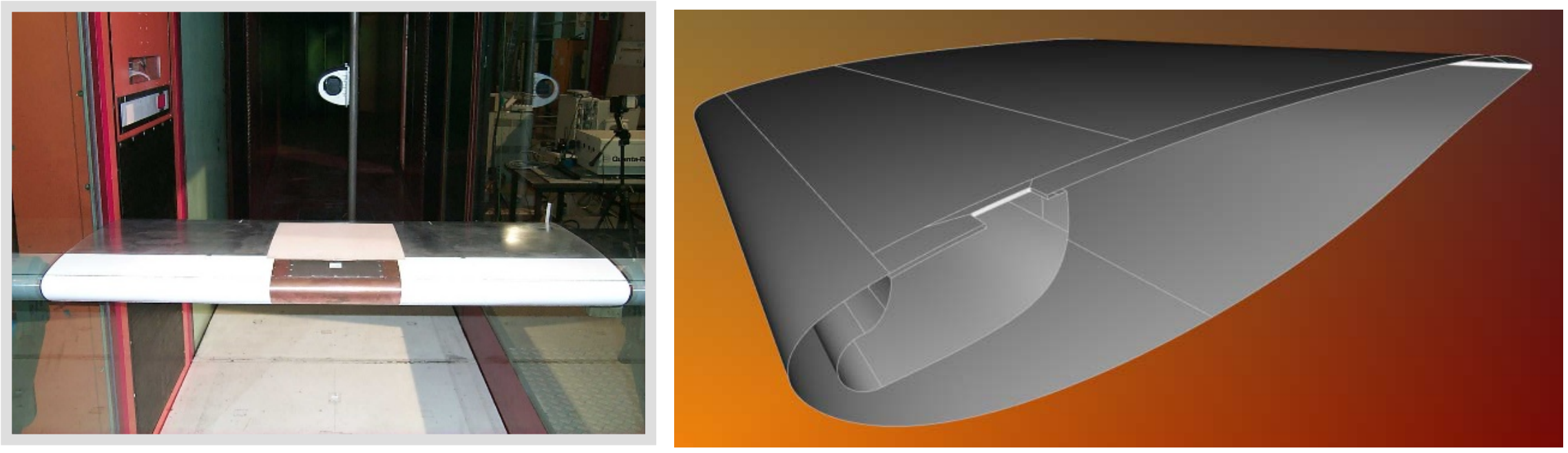

**Figure 7 - Wind tunnel setup and geometry of the MAEVA project**

We are interested here to recompute this test-case and the objective is to validate the specific injection boundary condition in an immersed boundary context. On part of the internal cavity surface, stagnation enthalpy and temperature are imposed in order to model the hot jet injection. The rest of surface is modelled with adiabatic walls with a Musker wall law. For this test-case, a quite coarse resolution has been chosen, with a step size of $10^{-3}$c in the vicinity of the square hole, $2.10^{-3}$c in the cavity, on the wing leading edge and in the jet wake and $8.10^{-3}$c elsewhere. The automatic IBM pre-processing leads to a Cartesian mesh of 63.5M points and 394 blocks.  A slice of the mesh with the resolution chosen on surfaces is presented on Figure 8. The computation has been performed on 56 Broadwell cores (Intel® Xeon® E5-2680v4 @ 2.4 GHz).

Figure 9 and Figure 10 show a comparison of the streamwise velocity flow field between experiments [1], previous elsA body-fitted mesh computation [2] and IBM solution. In the median plane, the flow is well captured by the IBM method, with an acceleration above the hole and a recirculation just aft of the hole. In cross sections, the solution features and the shape of the jet wake are correctly captured with both CFD methods but the flow field seems to dissipate too quickly according to the experiment, as the magnitude of the velocity is too low in the jet core.

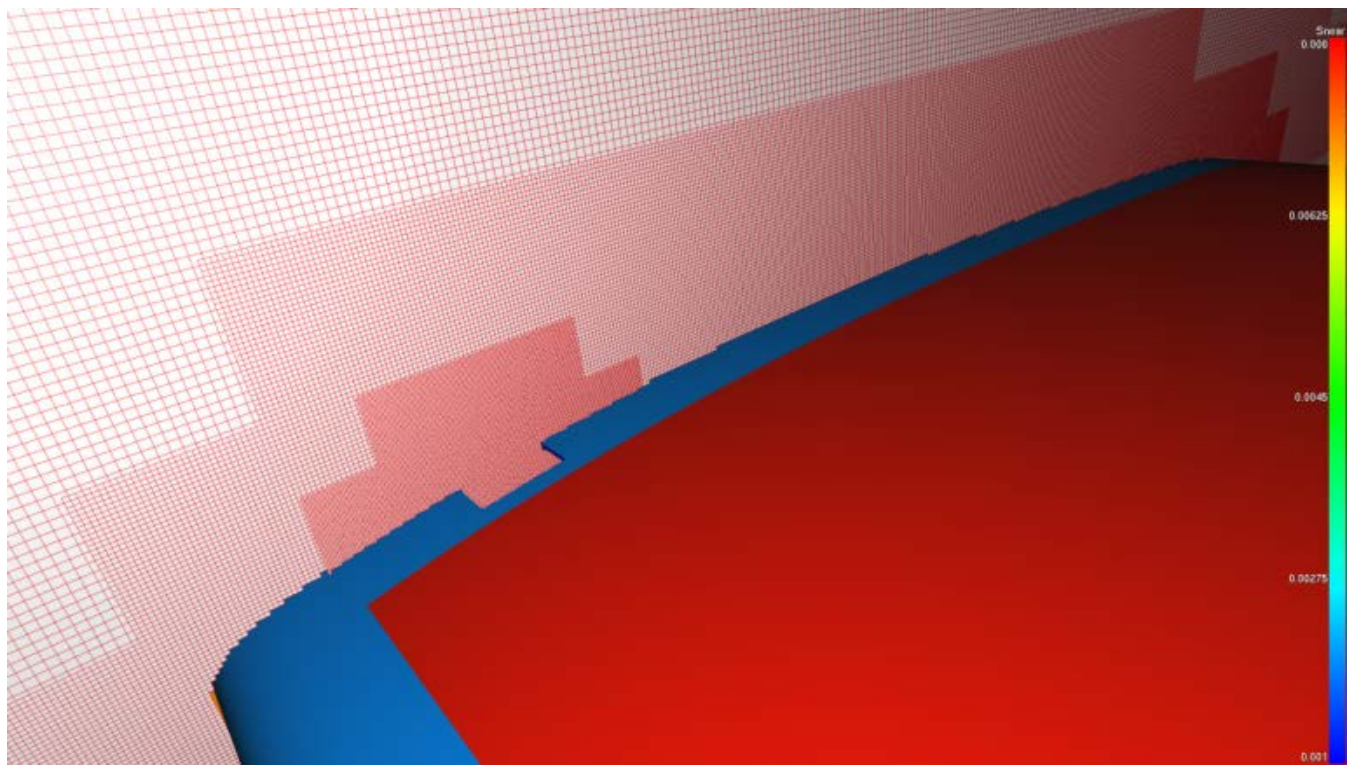

**Figure 8 - Slice of Cartesian mesh with local step size resolution**

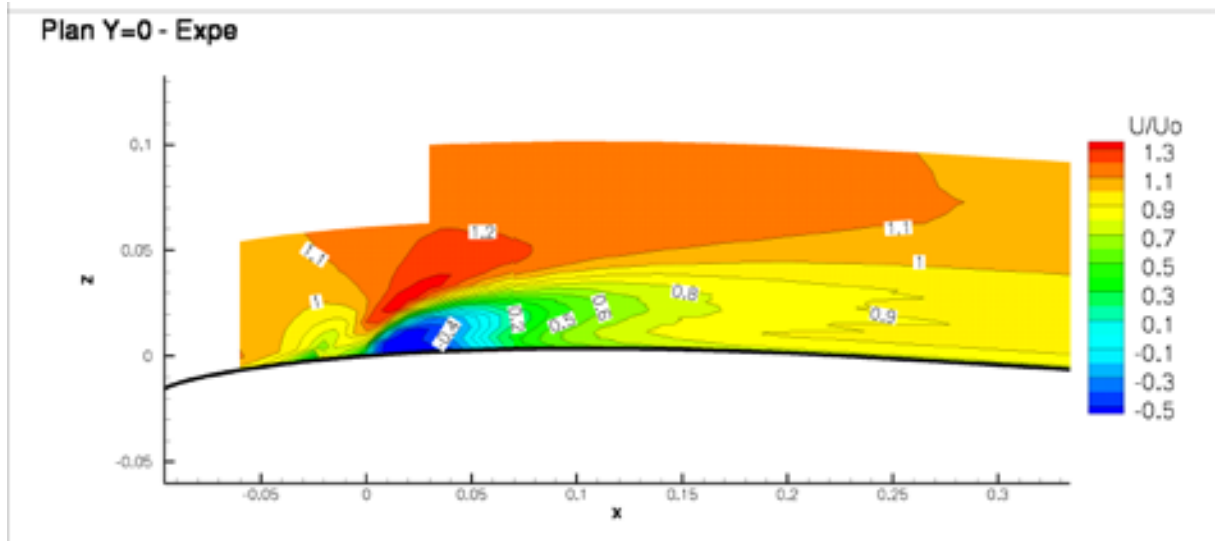

**(a) Experiment**

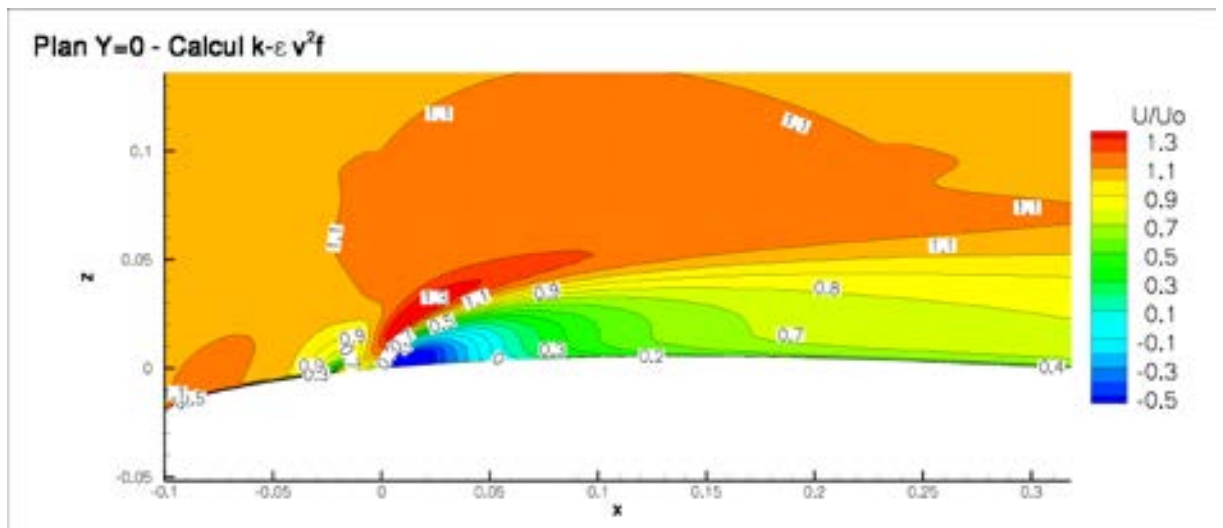

**(b) elsA computation with body-fitted mesh**

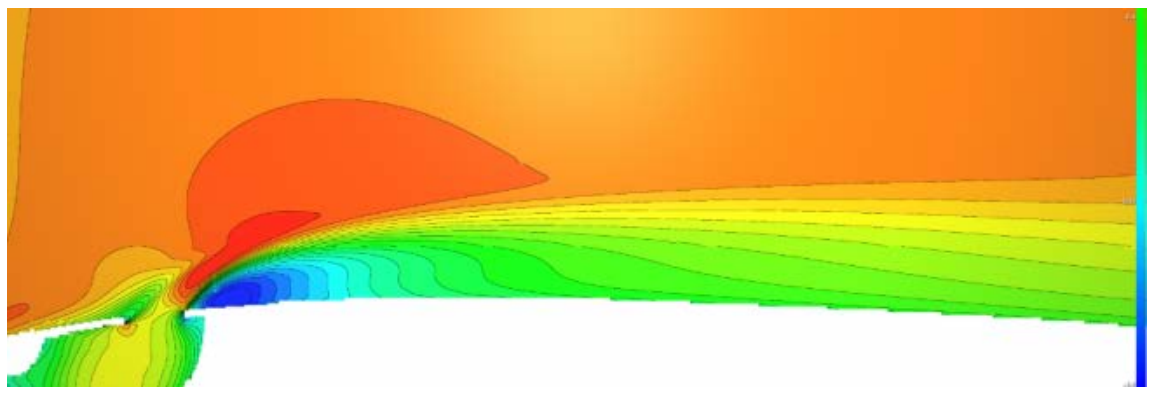

**(c) FAST computation with IBC method**

**Figure 9 - Comparison of streamwise velocity in the median plane**

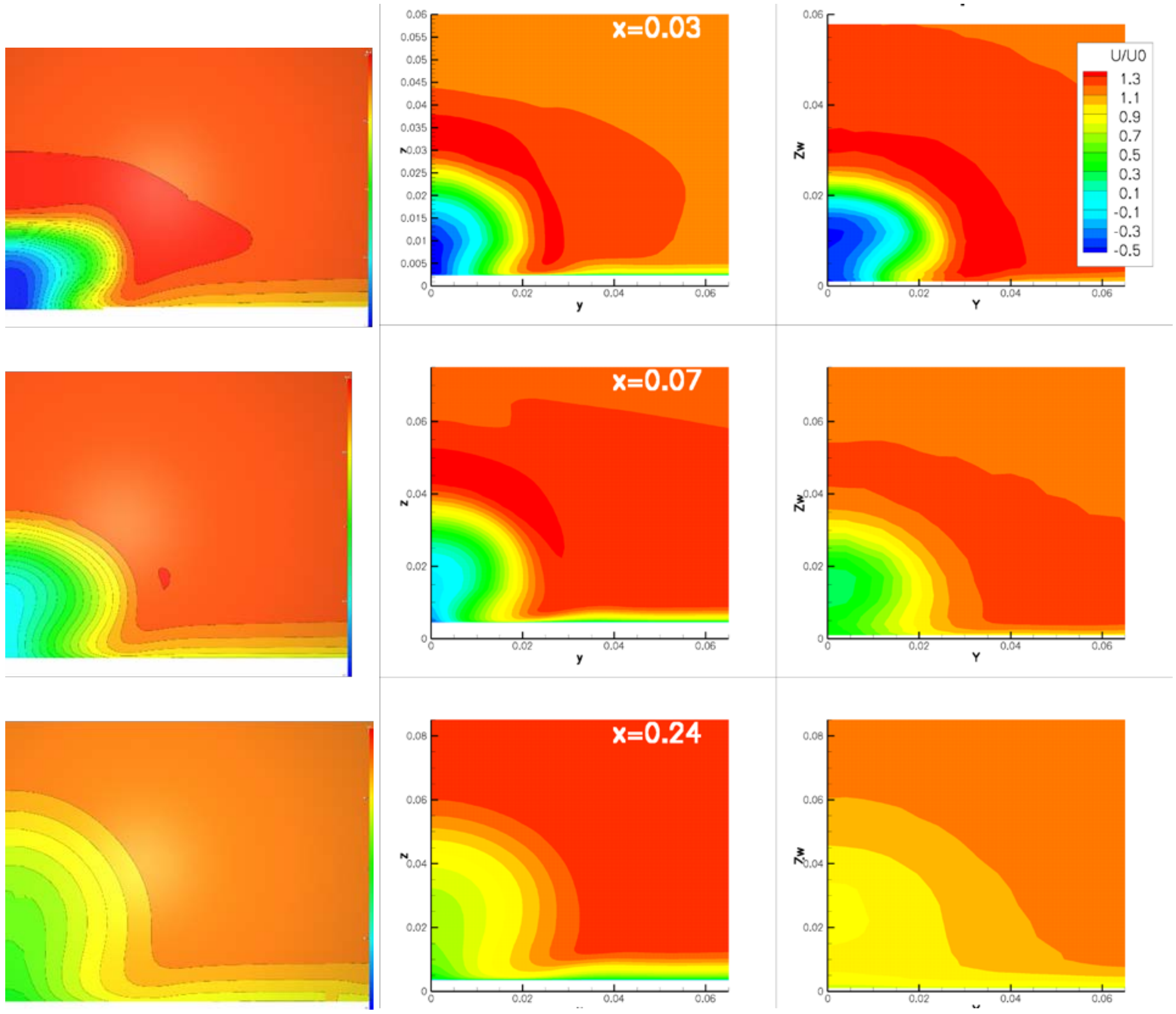

**(a) IBC method** **(b) elsA body-fitted** **(c) Experiment**

**Figure 10 - Comparison of streamwise velocity in the wake of the jet**

Figure 11 presents some profiles of the wall adiabatic efficiency $\eta_1$ in the median plane and in cross sections in the jet wake. The wall adiabatic efficiency is here defined by $\eta_1 = \frac{T_w - T_0}{T_j - T_0}$ where $T_w$, $T_j$ and $T_0$ are respectively the local wall temperature, the jet temperature in cavity and the far-field temperature. The elsA computation with body-fitted mesh and the FAST simulation with IBM approach are compared with the experimental data. In the IBM simulation, the efficiency is underestimated with respect to experiment in the symmetry plane, where the elsA solution is overestimated. In the cross sections, compared to classical body-fitted approach, the shape of the jet is better predicted by the IBM approach, as the hot temperature of the jet diffuses more in the spanwise direction. This is more consistent with the experimental data. The results should be improved by increasing the resolution of the IBM grid.

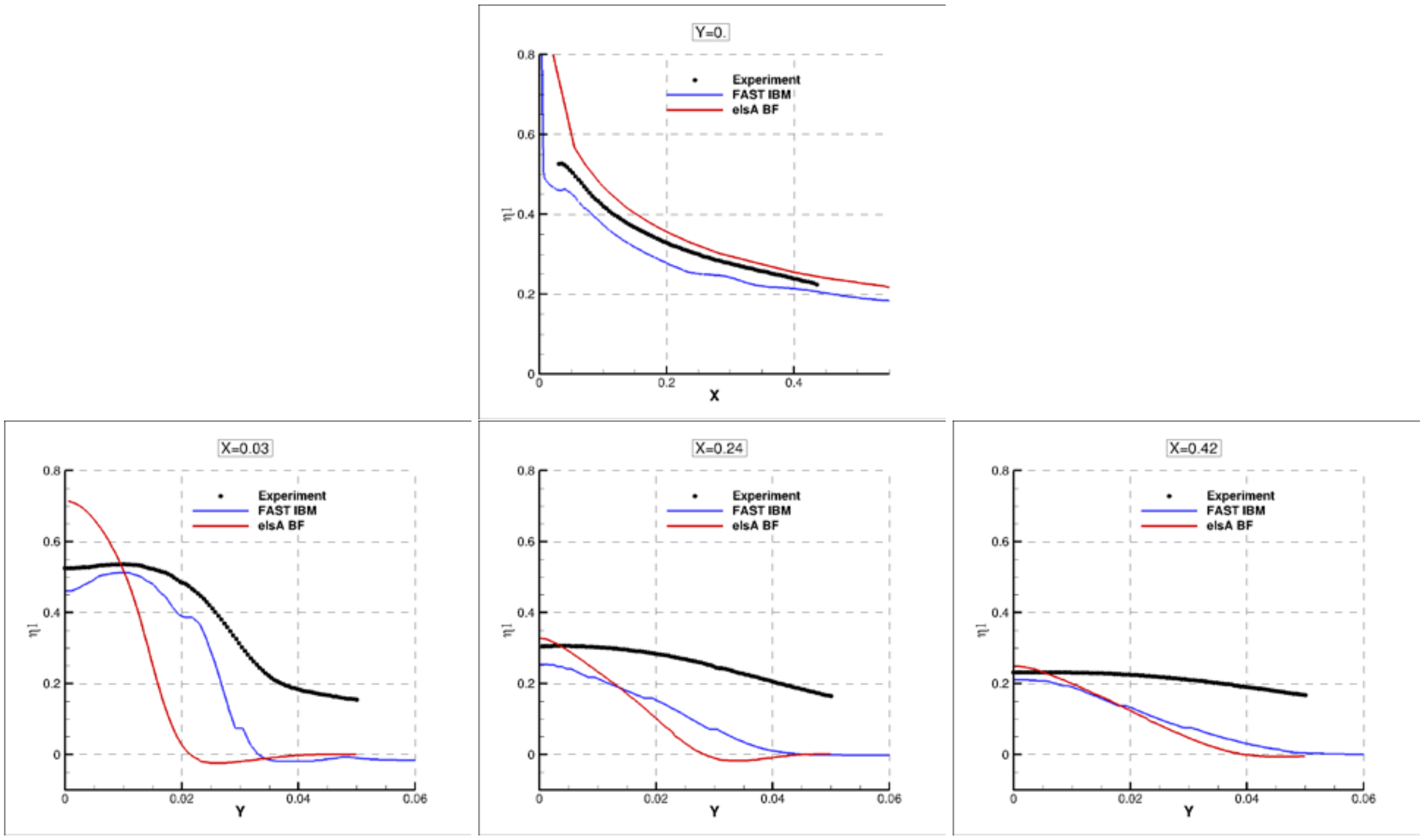

**Figure 11 - Comparison of wall adiabatic efficiency profiles (top: in the median plane, bottom: in streamwise sections)**

### V.C. 3D turbulent flow around a cylinder at Reynolds number $Re_d$=3900

At high Reynold number, three-dimensional effects and instabilities in the flow around a cylinder make the computation of this flow a challenging task for the IBM approach. For a Reynolds number $Re_d$=3900 (with respect to the cylinder diameter d and the free-stream velocity), literature provides several experimental and numerical results [3]. Here, the objective of this test-case is to validate the capability of the IBM method to capture the mean and unsteady flow characteristics. The free-stream Mach number is $M_\infty$=0.1. A large-eddy simulation is performed with a Musker wall law near the cylinder surface. The spatial scheme is a hybrid upwind/centered scheme based on the AUSM+P scheme with a sensor on flow regularity [17]. The time discretization is solved explicitly with a 3-steps Runge-Kutta scheme. The CPU cost of such a simulation can be compared with a computation with a body-fitted grid: even if the number of points of the IBM mesh is largely higher than the body-fitted one, it can be balanced with the dedicated Cartesian solver and the possibility to use a higher time step. Thus, the CPU gain is estimated to 24 in favor of the IBM approach.

A first grid (named "M1") is automatically generated around the cylinder: a Cartesian mesh is performed in two dimensions and then extruded in the spanwise direction. The length π*d of the cylinder is discretized with 48 points. The uniform resolution near the wall is equal to 0.006*d. Finally, the global 3D mesh contains 16.4M points and 117 blocks. A second mesh (named "M1 wake refined") is obtained by refining the M1 grid in the cylinder wake. Since the mesh generation tool in Cassiopee [15][16] enables to define refinement regions, a cone area is used here to refine the wake until 6*d. This second grid contains 40M points and 183 blocks. Finally, a third mesh (named "M2") has been generated by increasing the resolution (decreasing the cell size) near the cylinder. The uniform cell size near the wall is 0.003*d and the number of points in the spanwise direction is doubled (96 points). This third grid contains 89M points and 193 blocks. Figure 12 shows two-dimensional view of the three meshes. The computation on the finest grid has been performed on 56 Broadwell cores (Intel® Xeon® E5-2680v4 @ 2.4 GHz).

Figure 13 presents the instantaneous vorticity flow field in the wake of the cylinder obtained with the IBM M2 grid. At this high Reynolds number, the flow is unsteady, three-dimensional and leads to a Von Karman vortex alley. The vortices are rapidly dissipated due to the coarsening between the successive levels of the Cartesian mesh beyond a distance of 6*d aft of the cylinder.

The unsteady solution has been averaged during about 7*$T_c$, where $T_c$ is a period of the vortex cycle. It has also been averaged along the spanwise direction. The mean results are plotted on Figure 14, Figure 15 and Figure 16, compared to experiments [4][5]. On the first figure, the pressure coefficient on the cylinder shows the benefit of the M2 refinement near the wall: the IBM M2 solution fits the experimental values till 90° (upper/lower point of the cylinder) but the pressure is underestimated in the detached region (90°-180°). The wake refinement impinges slightly the wall results due to the recirculation of the turbulent structures. On Figure 15, the streamwise velocity is compared in the wake region on the center line. The solutions obtained with both M1 grids overestimated the length of the recirculation region, even if the wake refinement decreases it slightly. Once again, the major effect is brought with the M2 grid, whose solution is in better agreement with the experimental data. It is also confirmed with the streamwise velocity profiles in the cylinder wake on Figure 16. As before, the velocity profiles are in very good agreement with experiment till a distance of 2*d aft of the cylinder. Further and beyond a distance of 6*d, the streamwise velocity is underestimated due to the mesh coarsening.

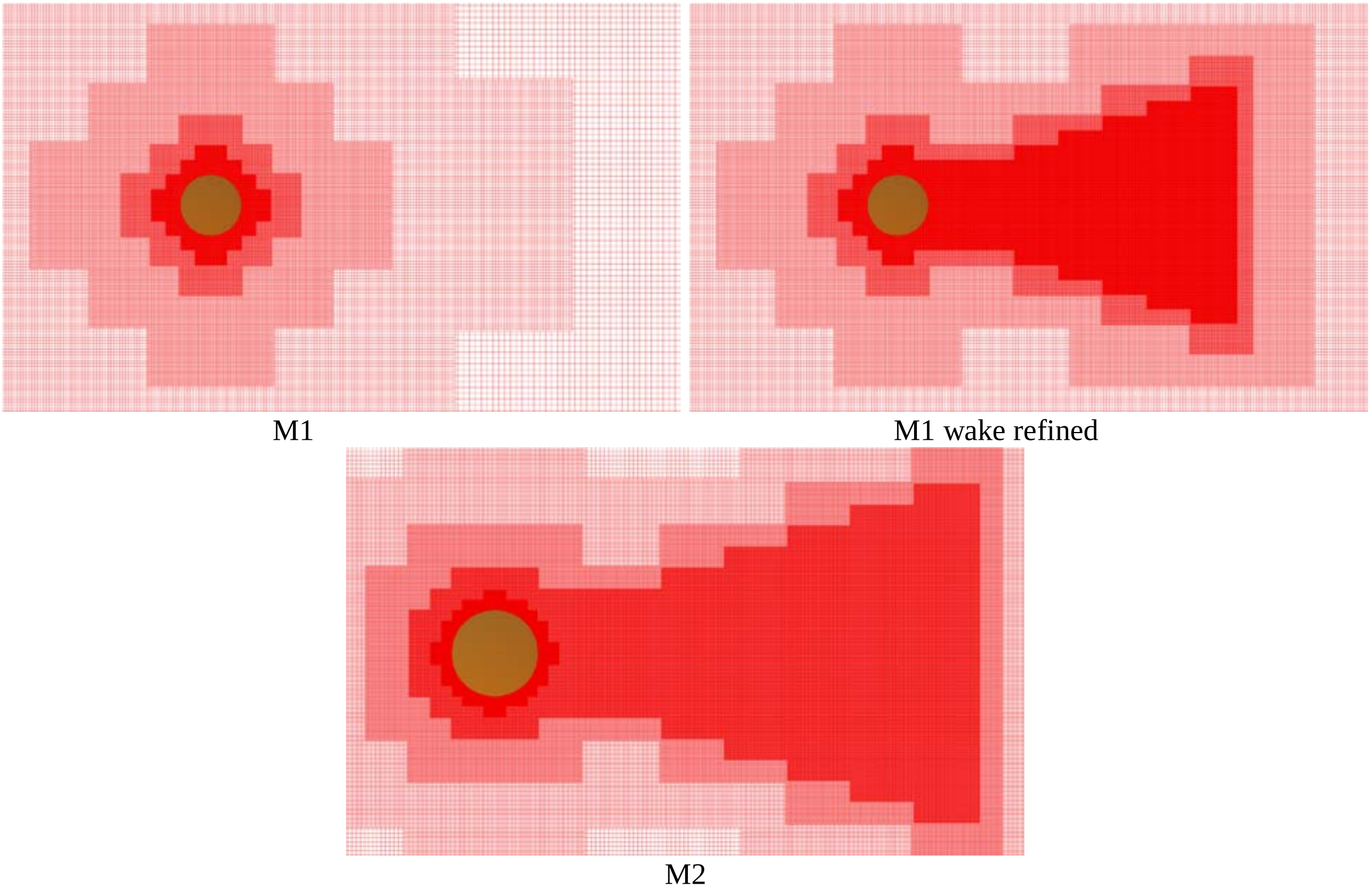

**Figure 12 - View of the three IBM meshes around the cylinder**

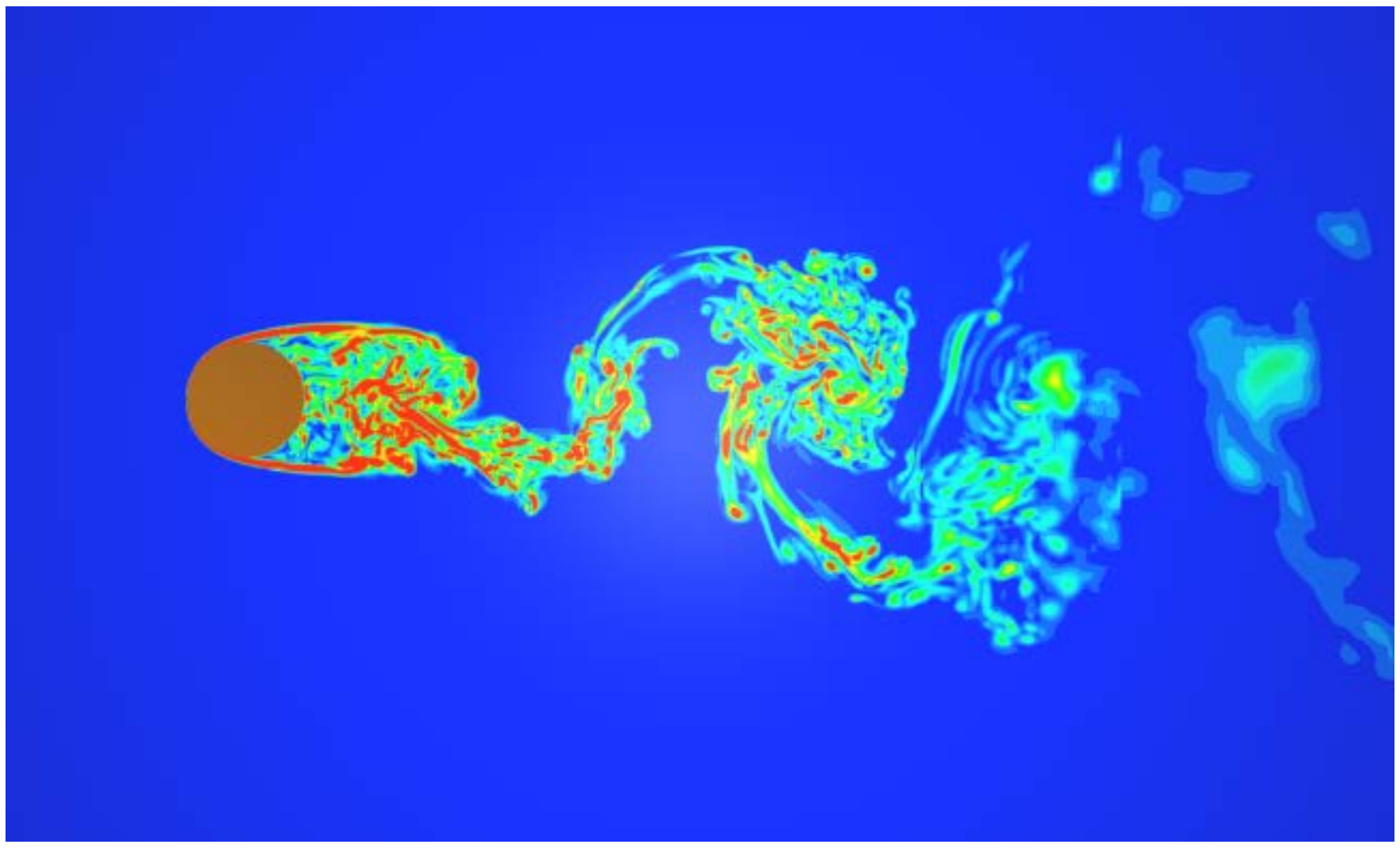

**Figure 13 - Instantaneous vorticity flowfield on IBM M2 grid (16 contours from $\omega d/U_\infty$=0.5 to $\omega d/U_\infty$=10.0)**

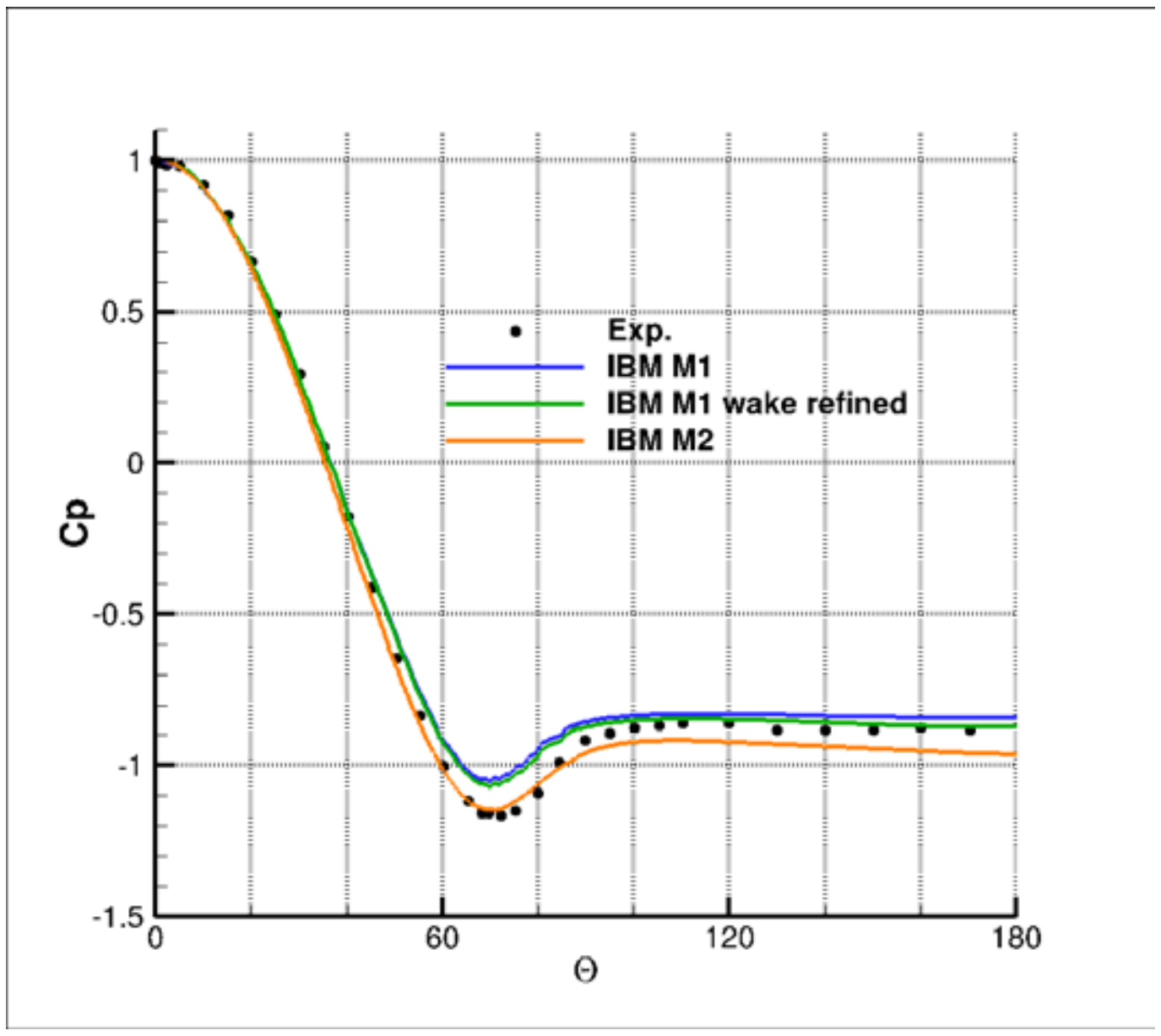

**Figure 14 - Pressure coefficient on the cylinder surface**

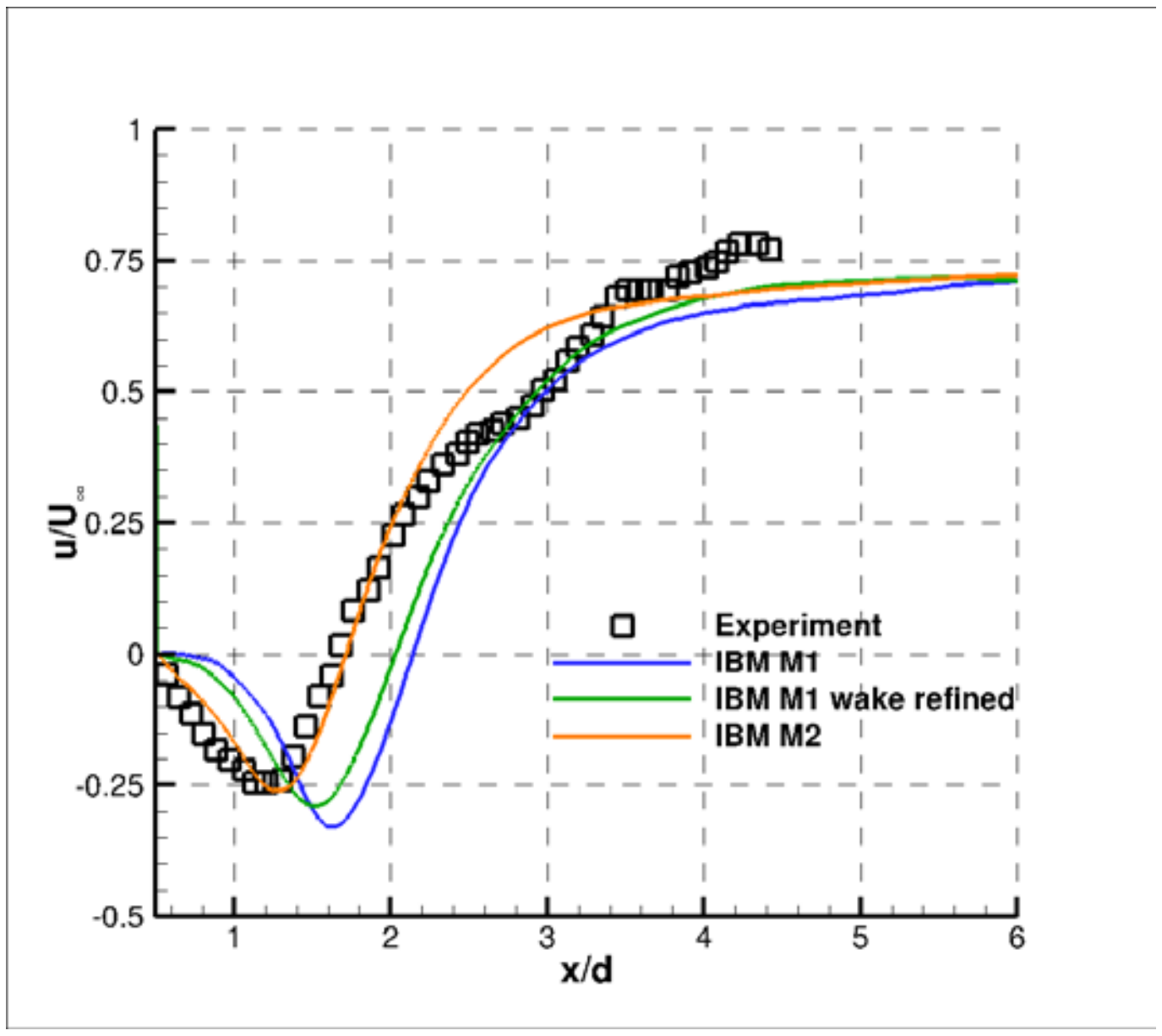

**Figure 15 - Streamwise velocity on the center line in the wake of the cylinder**

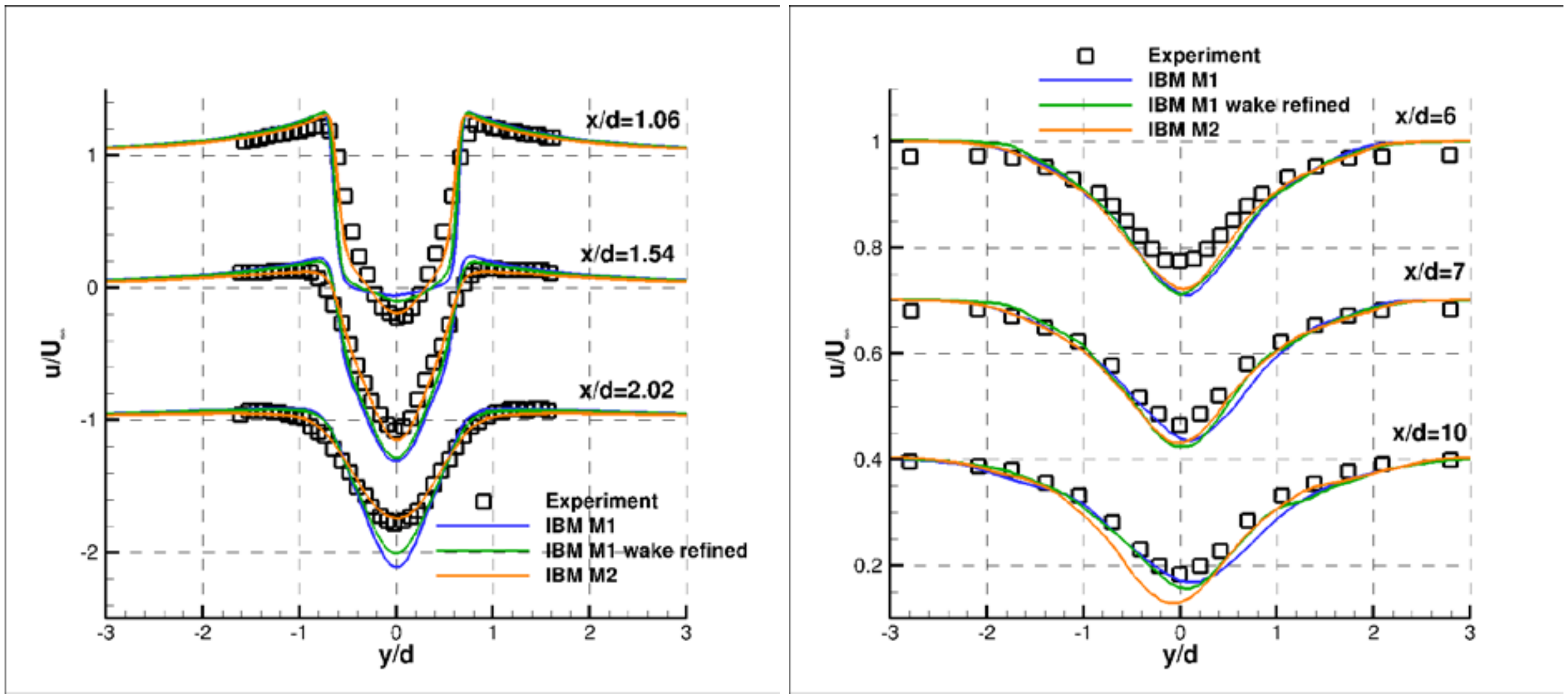

**Figure 16 - Streamwise velocity at 6 locations in the wake of the cylinder**

A spectral analysis of the unsteady flow field has been performed for the finest M2 grid. At first, Figure 17 presents the spectral power density calculated at the upper point of the cylinder (near the separation point) with respect to the Strouhal number. The graph exhibits a main peak at St=0.2, classical value for such a configuration. The peak at St=0.4 is not significant and the one at St=0.8 is more present. The second graph presents the spectral energy E11 ($m^2/s^2/Hz$) in the wake of the cylinder at x/d=1. As shown here with a main peak at St=0.4, at centerline, the flow oscillates at twice the Strouhal frequency due to velocity distribution in the Von Karman vortex street. At higher frequencies, the energy decreases with a -5/3 slope (represented in the graph by the dotted line).

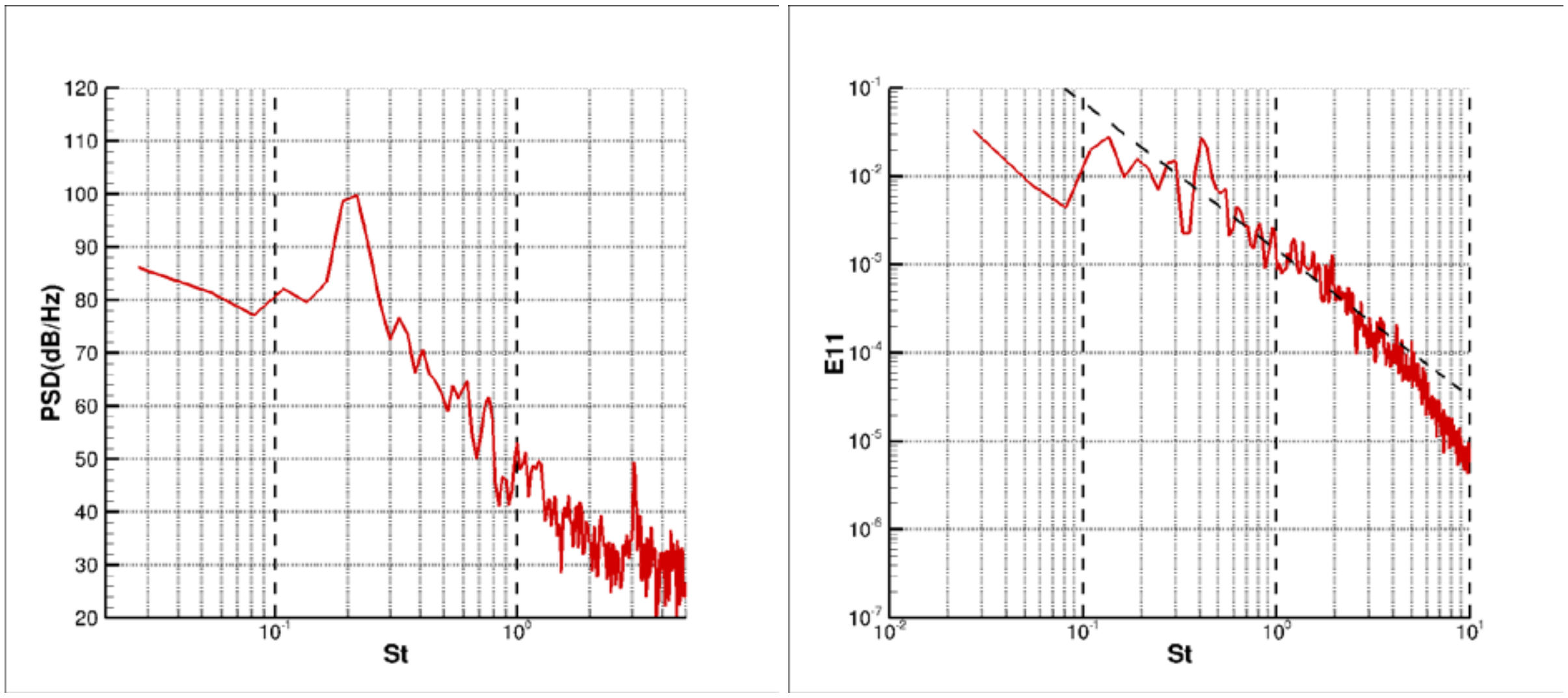

**Figure 17 - Spectral analysis of the flowfield: (left) spectral power density at cylinder upper point - (right) spectral energy at x/d=1.**

### V.D. 2D supersonic flow: the bow shock test-case

The bow shock test-case is one of the cases of the international workshop on high-order CFD methods [18]. In this workshop, this case is designed to isolate testing of the shock-capturing properties of schemes using the detached bow shock upstream of a two-dimensional blunt body in supersonic flow. This case is computationally expedient, being steady, two-dimensional, inviscid flow, with well-defined boundary conditions. In [19], the author shows that high-order schemes on unstructured grids are able to capture the shock location with very low pressure and enthalpy errors.

The geometry is a flat center section, with two constant radius sections top and bottom. The flat section is one unit length, and each radius is ½ unit length. While the flow is symmetric top and bottom, a full domain is computed to support potentially spurious behavior. The aft section of the body is not included to avoid developing an unsteady wake. The Mach number is $M_\infty$=4.

Here, our first objective is to validate the IBM approach for supersonic flows. The Cartesian grid is automatically generated around the body. The pre-processing allows to import refinement zones in order to better capture wakes for example. In this case, a refinement region is defined by a circle of radius 3.5 unit length around the blunt body. A uniform cell size of 0.01 unit length is imposed near the wall and inside the circle.

One of the advantages of our method is to propose the adaptation of the Cartesian grid with respect to a given criterion: flow characteristic, numerical error… As based on an Octree grid, an indicator can be computed from a first solution. The pre-processing tool can then adapt the Octree grid with respect to it in order to adapt the final mesh. In the present test-case, the Mach number difference is chosen in order to adapt the grid around the bow shock. Figure 18 presents both meshes: the original one with the circular refinement region contains 0.73M points and 47 blocks, the adapted one contains 1.27M points and 190 blocks. In the adapted mesh, the minimum cell size in the vicinity of the shock is $5.10^{-4}$ unit length. It can be noticed that some regions have been coarsened with respect to the original grid.

Computations have been performed with the FAST solver and a 2$^{nd}$-order Roe scheme ("minmod" limiter). Figure 19 shows the flow field through iso-contours of Mach number. With the adapted grid, the shock is obviously thinner. Some small oscillations can be noticed near the shock, due to the spatial scheme. For steady inviscid flow, the total enthalpy, $H = (\rho E + p)/\rho$, is constant, where ρE is the total energy. The error in this quantity provides a first quantifiable measure of the quality of the computed solution of the general Euler equations (as opposed to schemes which specifically optimize for steady, inviscid flow and enforce H = const.). Along the stagnation streamline, the stagnation pressure on the cylinder surface is predicted by the Rayleigh-Pitot formula,

$$\frac{p_{02}}{p_1} = \frac{1-\gamma+2\gamma M_1^2}{\gamma+1} * \left(\frac{(\gamma+1)^2 M_1^2}{4\gamma M_1^2 - 2(\gamma-1)}\right)^{\frac{\gamma}{\gamma-1}}$$

where subscript 1 refers to conditions upstream of the shock and 2 to the stagnation point.

Table 2 gives the wall pressure ratio results. The adapted mesh improves this value, by dividing by 10 the error with respect to the theory.

**Table 2 – Comparison of pressure ratios**

| | Theory | Original grid | Adapted grid |
|---|---|---|---|
| $\frac{p_{02}}{p_1}$ | 21.068081 | 21.042294 | 21.065672 |

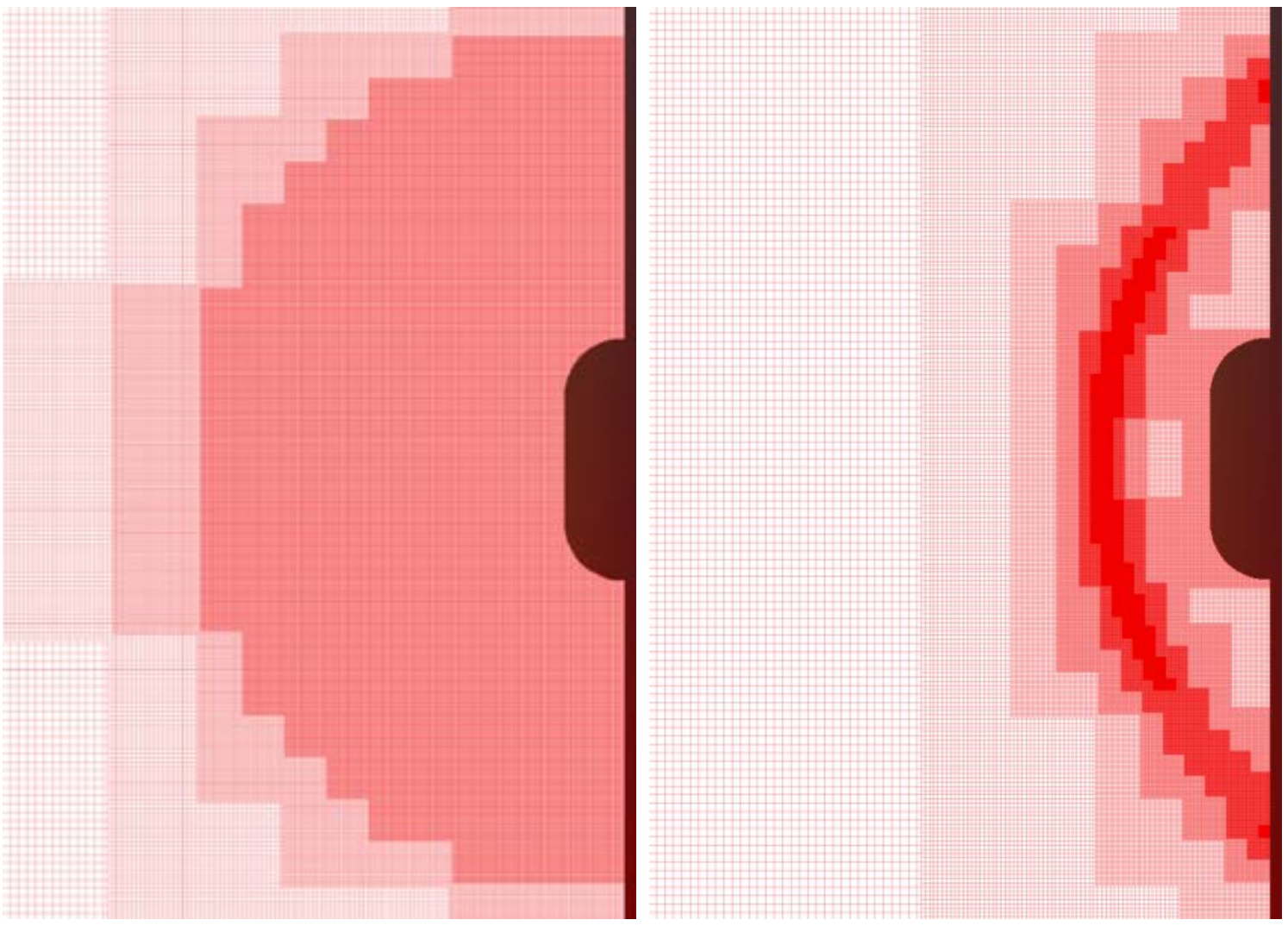

**Figure 18 - 2D Cartesian grids around a blunt body – Original (left) and adapted (right) meshes**

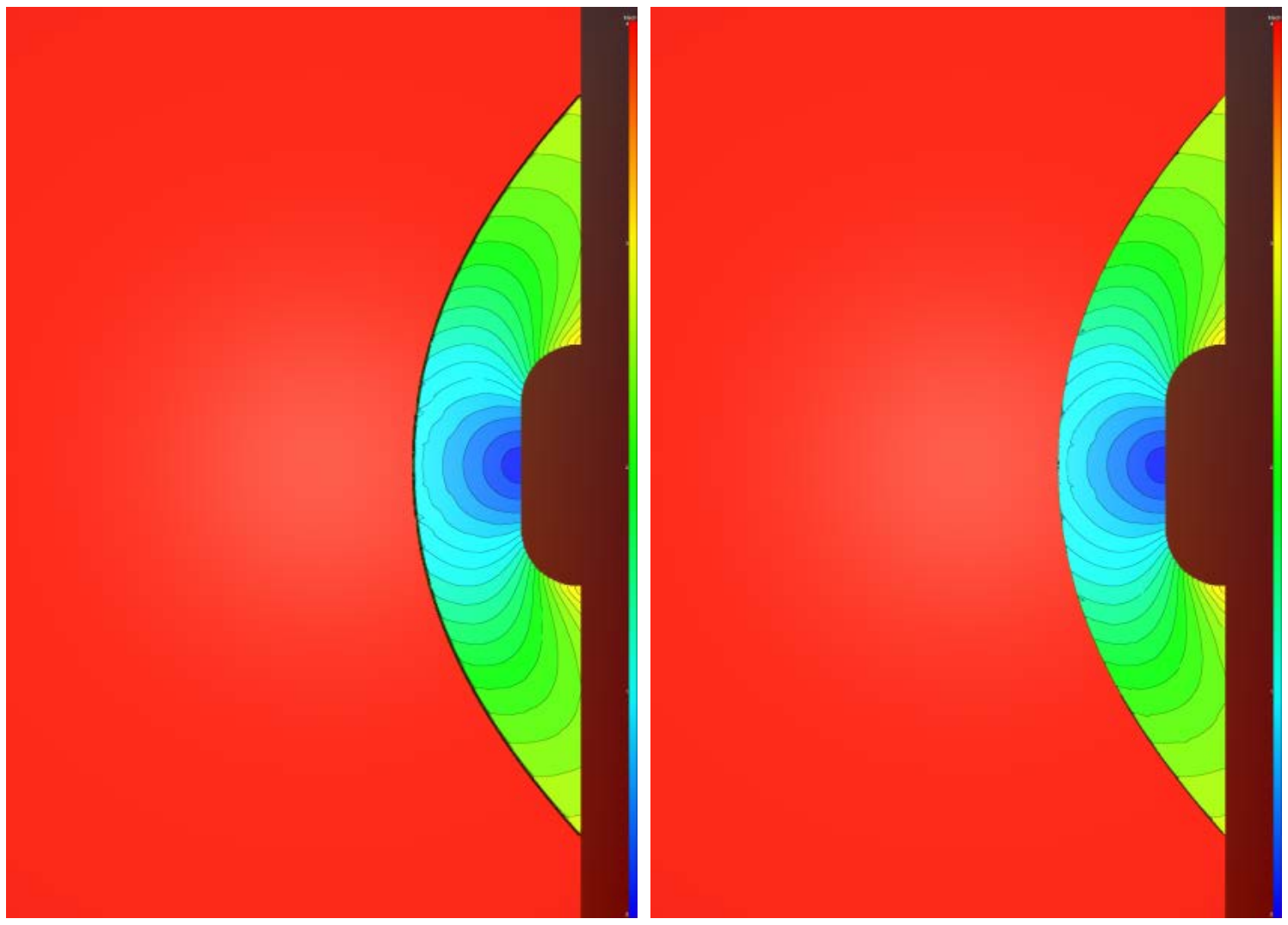

**Figure 19 - Isocontours of Mach number around the blunt body**

## VI. Conclusion

This papers demonstrates that a method based on an automatically generated Cartesian mesh and where body are taken into account by immersed boundary conditions provide accurate results for well-known compressible test-cases for aircraft aerodynamics. For various flow regimes (from subsonic to hypersonic), the present IBM approach in combination with a Cartesian mesh that is automatically generated and a dedicated Cartesian HPC solver (FAST solver) enable to obtain very satisfactory results within a limited time. It can be all the more efficient for unsteady LES simulations, as presented in the present paper for the cylinder test-case. At the present time, the method allows some immersed boundary conditions like walls, injection, imposed pressure but will be enriched in a close future. The ease of pre-processing and the solver efficiency allows the engineer to obtain a reliable CFD solution in a short time (within a day for steady computations).